\documentclass[a4paper]{jpconf}
\usepackage{graphicx}

\begin{document}

\title{Dissipative hydrodynamics and heavy ion collisions}

\author{\bf A. K. Chaudhuri}
\address{Variable Energy Cyclotron Centre,
1/AF,Bidhan Nagar, Kolkata - 700 064}
\ead{akc@veccal.ernet.in}

\begin{abstract}
Space-time evolution and subsequent particle production from    minimally viscous ($\eta/s$=0.08) QGP fluid  is studied using the 2nd order   Israel-Stewart's theory of dissipative relativistic fluid. Compared to ideal fluid, energy density or temperature evolves slowly in viscous dynamics. Particle yield
  at high $p_T$ is increased. Elliptic flow on the other hand decreases in viscous dynamics.   
Minimally viscous QGP fluid found to be consistent with a large number of experimental data.
\end{abstract}

\section{Introduction} \label{sec1}
%
 
Success of {\em ideal} hydrodynamics \cite{QGP3},
in explaining bulk of the data in Au+Au collisions at RHIC,   has led to a paradigm     that in Au+Au collisions, a nearly perfect fluid is created.  
However, the paradigm of {\em perfect fluid}, produced in Au+Au collisions at RHIC, need to be clarified.  
Experimental data do show deviations from ideal behavior; at large $p_T > 1.5$ GeV, in peripheral collisions or at forward rapidity  \cite{Heinz:2004ar}, presumably due to increasingly important role of dissipative effects. ADS/CFT correspondence \cite{Policastro:2001yc} also suggests that in a strongly coupled matter, the shear viscosity to entropy ratio  is bounded from the lower side, $\eta/s \leq 1/4\pi$. At the minimum, $\eta/s=1/4\pi \approx 0.08$ for QGP fluid.
Hydrodynamics is applicable only when, $\eta/s << T\tau$ \cite{Teaney:2003kp}. 
In Au+Au collisions at RHIC,   at the early time, $\tau_i$=0.6-1.0 fm,  temperature is $T_i\approx $300-350 MeV, which limit the 
viscosity to $\eta/s <<$ 0.9-1.8. ADS/CFT bound is smaller by factor of 10-20 from the limiting viscosity.

Though the theories of dissipative hydrodynamics \cite{Eckart,LL63,IS79} 
has been known for more than 30 years , significant progress toward its numerical implementation  has only been made very
recently 
\cite{Teaney:2003kp,MR04,Koide:2007kw,Chaudhuri:2005ea,Heinz:2005bw,asis1st,asis2nd,Romatschke:2007mq,Song:2007fn,Dusling:2007gi}. Several groups have solved the equations for causal hydrodynamics.
The program is still ongoing and consensus is not reached between different groups.   In the following, we discus some aspect of viscous dynamics and associated results.
 
\section{Causal dissipative hydrodynamics}

We consider  QGP fluid in the central rapidity region, with 
zero net baryon density and chemical potential, $n_B=0,\ \mu_B=0$. We neglect 
the effects of heat conduction ($\mu_B=0$) and bulk viscosity   and account only for the shear viscosity. We work in the 
Landau-Lifshitz energy frame.
The energy-momentum tensor, including the shear stress  
tensor $\pi^{\mu\nu}$, is written as,
\begin{equation} \label{eq1}
T^{\mu\nu} = (\varepsilon+p)u^\mu u^\nu - p g^{\mu\nu} + \pi^{\mu\nu},
\end{equation}
where  $\varepsilon$ is the energy density, $p$ is the
hydrostatic pressure, and $u$ is the hydrodynamic 4-velocity,
normalized as $u^\mu u_\mu=1$. $T^{\mu\nu}$ satisfies the 
energy-momentum conservation law, 
\begin{equation} \label{eq2}
\partial_\mu T^{\mu\nu} = 0.
\end{equation}

In the Israel-Stewart's 2nd order theory of dissipative fluids \cite{IS79}, the
dissipative fluxes are treated as extended thermodynamic variables
which satisfy   relaxation equations. The   relaxation
equation for the shear stress tensor read
\begin{equation} \label{eq3} 
D\pi^{\mu\nu}=\frac{1}{\tau_\pi}(\pi^{\mu\nu} - 
2\eta\nabla^{\left\langle\mu\right.} u^{\left.\nu\right\rangle}),
\end{equation}
where  $D=u^\mu\partial_\mu$ is the convective time derivative,  
$\eta$ is the shear viscosity coefficient and $\tau_\pi$ is the relaxation time. 
$\nabla^{\left\langle\mu\right.} u^{\left.\nu\right\rangle}$ is a  symmetric, traceless tensor, 

\begin{equation} \label{eq4}
\nabla^{\left\langle\mu\right.} u^{\left.\nu\right\rangle}
  = \frac{1}{2}[ \nabla^\mu u^\nu+\nabla^\nu u^\mu] -\frac{1}{3}
    \Delta^{\mu\nu} \partial_\sigma u^\sigma,
\end{equation}
where $\nabla^\mu = \partial^\mu-u^\mu D$ is the transverse gradient operator and $\Delta^{\mu\nu}=g^{\mu\nu}-u^\mu  u^\nu$ is the
projector orthogonal to the flow velocity $u^\mu$.
In the time scale $\tau_\pi$, viscous pressure relaxes to 1st order value,
 $2\eta\nabla^{\left\langle\mu\right.} u^{\left.\nu\right\rangle}$ \cite{Eckart,LL63}. 
 The viscous pressure tensor $\pi^{\mu\nu}$ is symmetric ($\pi^{\mu\nu} = \pi^{\nu\mu}$),   traceless    ($\pi^\mu_\mu =0$) and transverse to the hydrodynamic 4-velocity, $u_\mu \pi^{\mu\nu}=0$. It has 5 independent components.  There are 10 unknowns ($\varepsilon$, $p$, three components of the hydrodynamic velocity  $u$, and 5 viscous pressure components) and 9 equations (4 energy-momentum conservation equations and 5 transport equations for the independent components of $\pi^{\mu\nu}$). The set of equations is closed by the equation of state $p=p(\varepsilon)$.
We note that presently there is disagreement about the form of the relaxation equation as given in Eq.\ref{eq3}. 
Some authors \cite{Romatschke:2007mq,Song:2007fn} have solved the relaxation
equation containing an extra term $R=[u^\mu\pi^{\nu\lambda}+u^\nu\pi^{\nu\lambda}]Du_\lambda$.
The term ensures that throughout the evolution shear stress tensor remains traceless and transverse to fluid velocity. Israel-Stewart 
developed the theory on gradient expansion of entropy density. Gradients of equilibrium thermodynamical variables  are
assumed to be small. Since dissipative flows are small, terms like  $\pi^{\mu\nu}Du_\nu$ is neglected. Consequently, Eq.\ref{eq3} is restricted to situations where gradient of velocity is small.  In the present paper, we have limited our study to minimally viscous fluid ($\eta/s$=0.08). It will be shown later that for minimally viscous fluid, contribution of the term R is small and fluid evolution  is hardly changed whether the term R is included or not in the relaxation equation.  
\begin{figure}[t]
\begin{minipage}{11pc} 
\includegraphics[bb= 40 278 507 755  
    ,width=1.0\linewidth,clip]{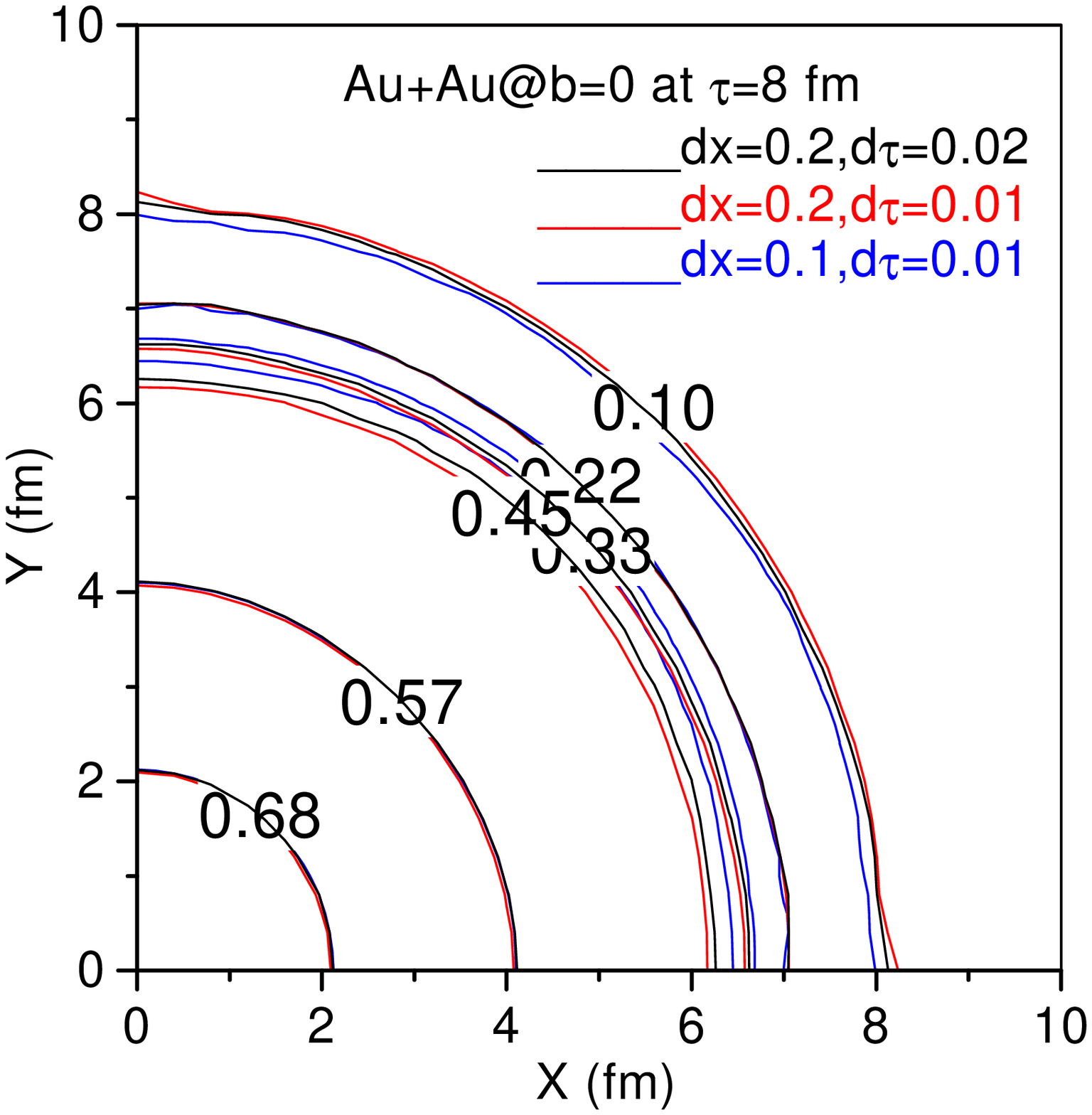}
\end{minipage} \hspace{1pc}
\begin{minipage}{11pc} 
\includegraphics[bb=44 289 526 770  
  ,width=1.0\linewidth,clip]{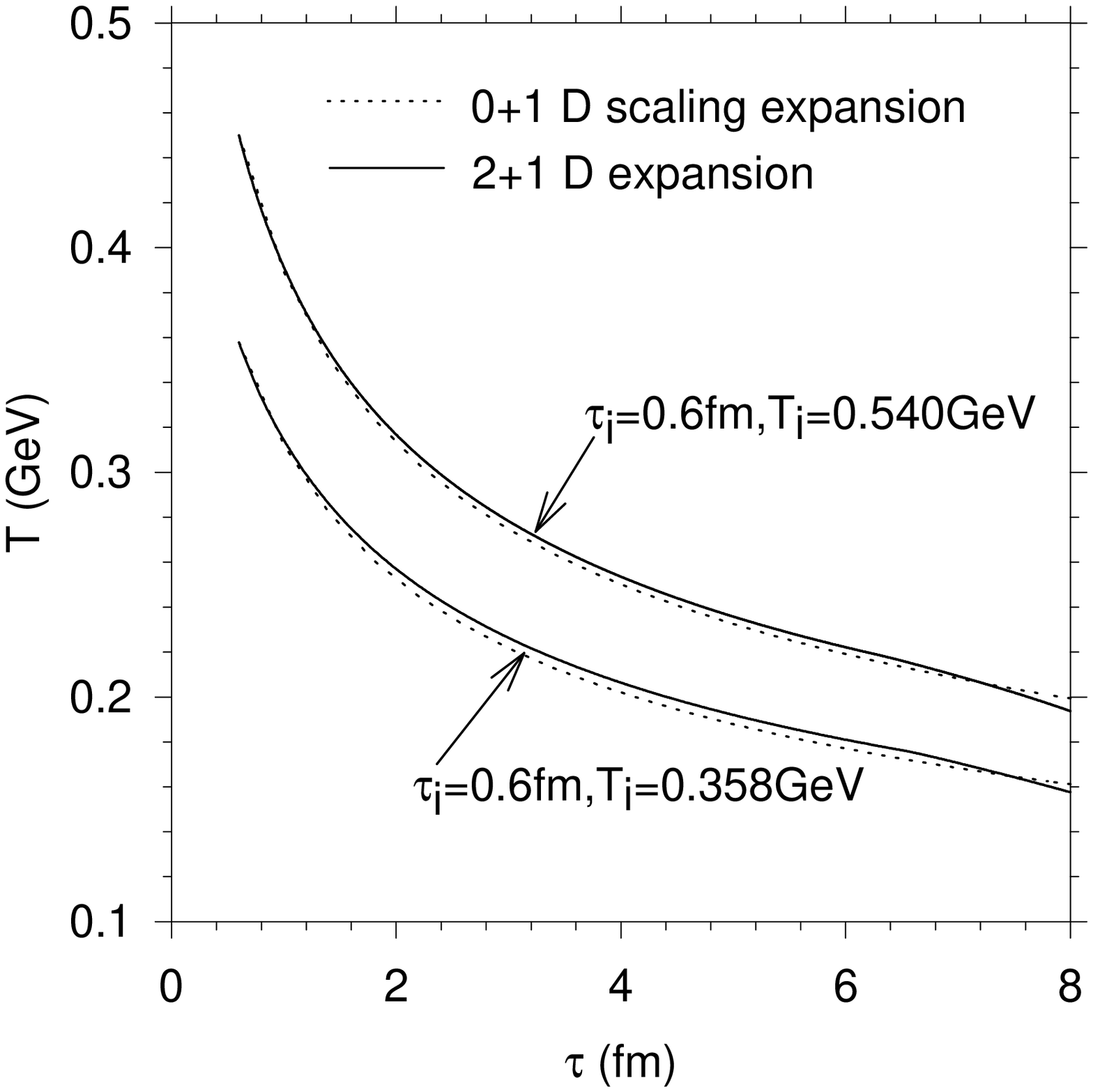}
\end{minipage}\hspace{1pc}
\begin{minipage}{11pc}
\includegraphics[bb=47 246 527 736
    ,width=1.0\linewidth,clip]{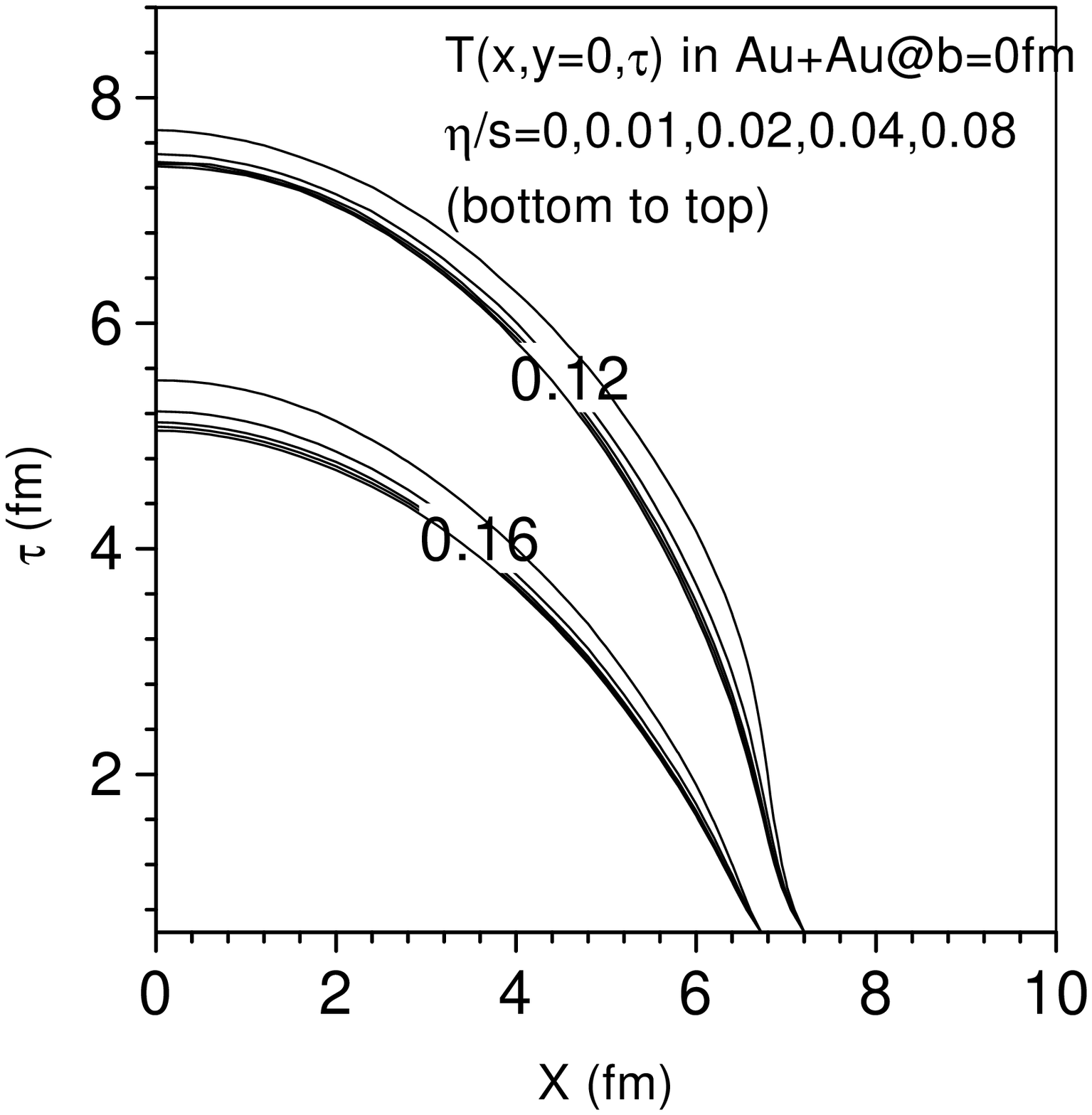}
\end{minipage}
\caption{\label{F1} (color on line). For testing the computer code AZHYDRO-KOLKATA, we have considered fluid evolution in QGP phase only.
In the left panel, contour plot of energy density show that the evolution 
is stable with respect to change in integration step-lengths, maintain any given symmetry and do not lead to unphysical maxima or minima.  The
middle panel shows that the fluid at the centre is least affected by the transverse expansion, and closely follow the 0+1 dimensional causal flow. 
In both left and middle panel viscosity to entropy ratio is $\eta/s$=0.08.
 The right panel shows that ideal hydrodynamic results are recovered when viscosity gradually reduces to zero.}
\end{figure}

\begin{figure}[t]
\begin{minipage}{16pc}
\includegraphics[width=16pc]{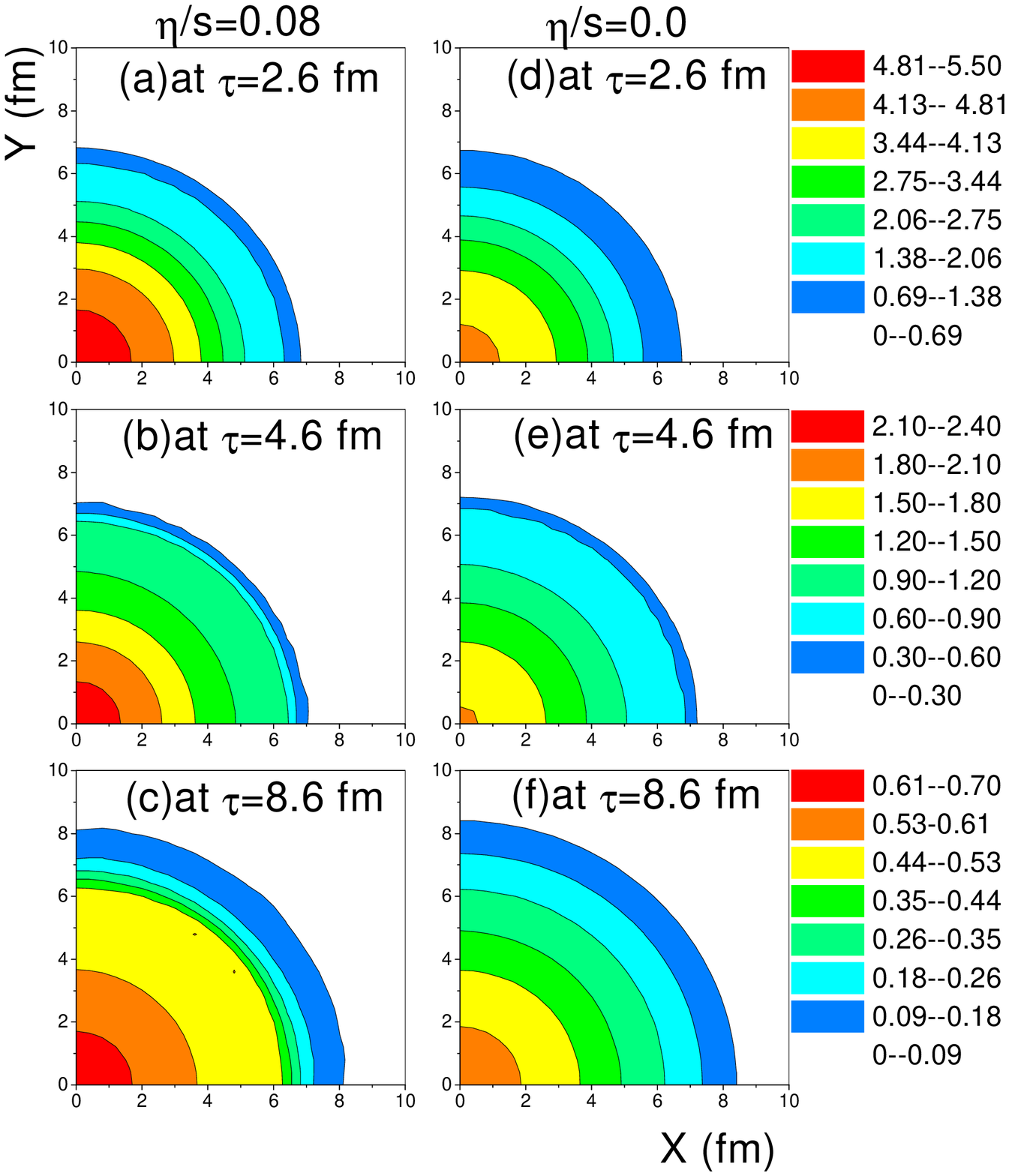}
\caption{\label{F2}Evolution of energy density in ideal
and minimally viscous  flow in b=0 Au+Au collision.}
\end{minipage}\hspace{4pc}%
\begin{minipage}{16pc}
\includegraphics[width=13.5pc]{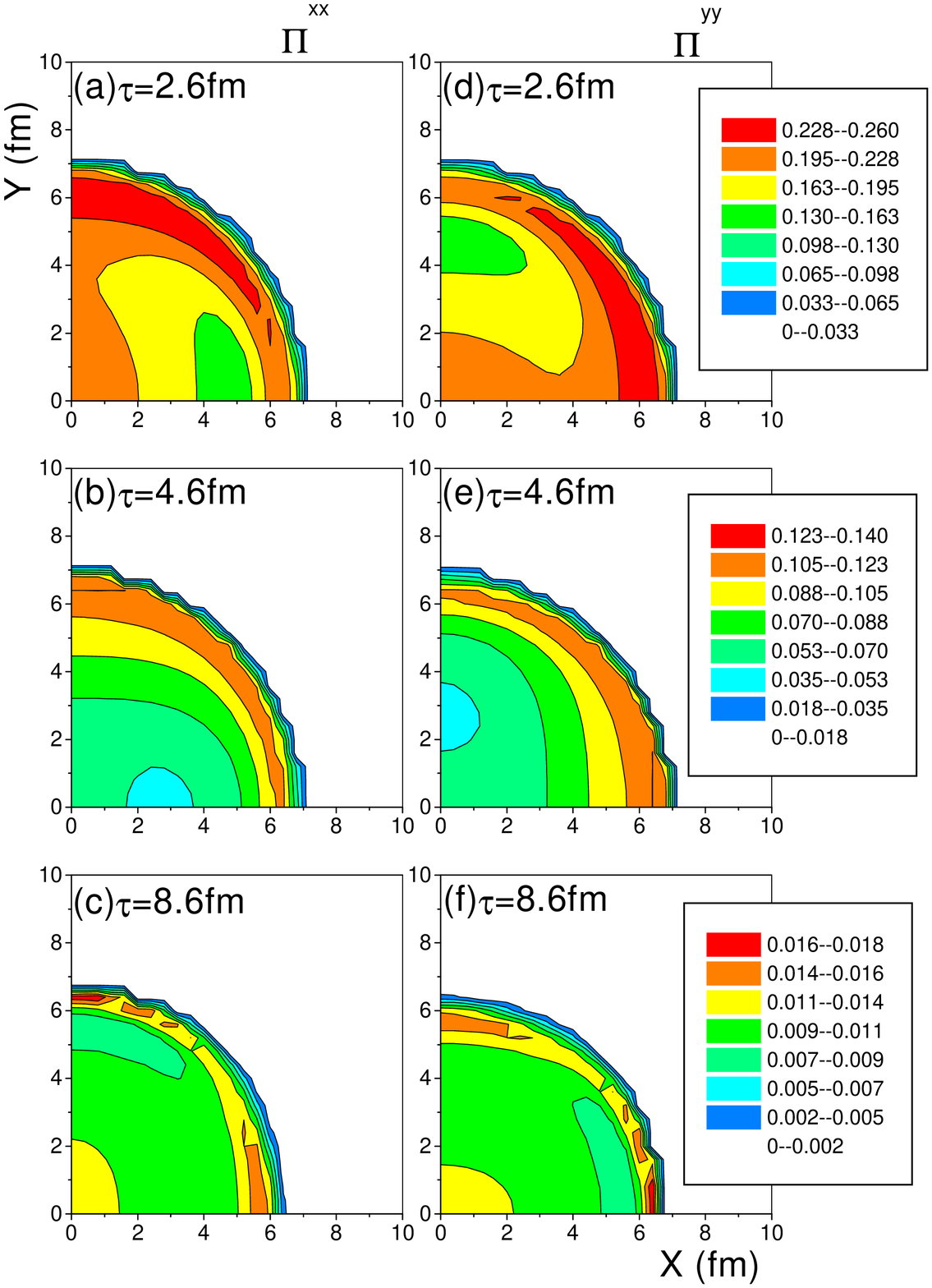}
\caption{\label{F3}Evolution of stress tensor $\pi^{xx}$ and $\pi^{yy}$ in b=0 Au+Au collision. $\eta/s$=0.08.}
\end{minipage} 
\end{figure}

Assuming   longitudinal boost-invariance, we solve Eqs.\ref{eq2} and \ref{eq3} in
 ($\tau=\sqrt{t^2-z^2}$,x,y,$\eta=\frac{1}{2} \ln \frac{t+z}{t-z}$) coordinates. The assumption reduces the number of unknowns to 6 ($\varepsilon$, two components of $u$ and three  components of shear stress tensor. We choose $\pi^{xx}$, $\pi^{yy}$ and $\pi^{xy}$ as the independent shear stress tensor components. There can be other choices also  \cite{Romatschke:2007mq,Song:2007fn}. 
Six partial differential equations  are simultaneously solved with the code "AZHYDRO-KOLKATA" developed at the Cyclotron Centre, Kolkata.
We use the standard initial conditions and equation of state EOS-Q (incorporating 1st order phase transition at $T_c$=164 MeV), described in \cite{QGP3}.
At initial time $\tau_i$=0.6 fm, the QGP fluid was initialized with central entropy density $S_{ini}=110 fm^{-3}$, with a Glauber model transverse density profile. It corresponds to peak energy density $\sim 35 GeV/fm^3$. Initial fluid velocity is zero.  
 In viscous hydrodynamics, 
additionally, one has to initialize the shear stress-tensor. 
Viscous effects are enhanced if initially $\pi^{\mu\nu}$ is
non-zero rather than zero. We choose to maximise the effect of viscosity. Instead of choosing any arbitrary non-zero value, we initialise shear stress-tensor to boost-invariant value, $\pi^{xx}=\pi^{yy}=2\eta/3\tau_i$, $\pi^{xy}=0$.
For the relaxation time, we use the kinetic theory approximation for a Boltzmann gas, $\tau_\pi=\frac{3\eta}{2p}\approx \frac{6\eta}{sT}$. For the viscosity, we have used the ADS/CFT lower bound $\eta/s=0.08$. Throughout the evolution, bound on  the shear viscosity to entropy ratio is maintained.

\begin{figure}[t]
\includegraphics[width=16pc]{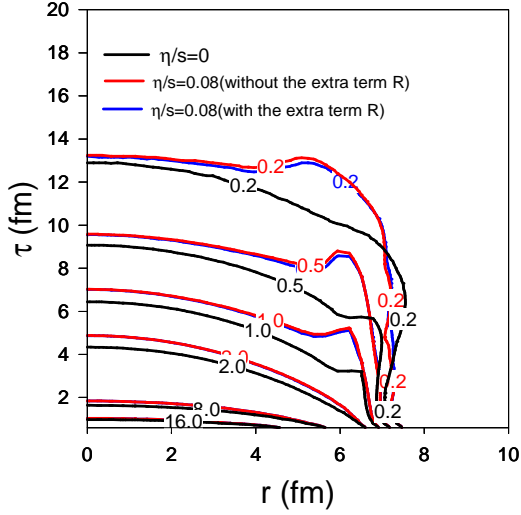}\hspace{2pc}%
\begin{minipage}[b]{16pc}
\caption{\label{F4}
(color online) Evolution of energy density in a b=0 fm Au+Au collision. The black lines are constant energy density contours in $r(=\sqrt{x^2+y^2})-\tau$ plane in ideal dynamics.
The red lines are for minimally ($\eta/s$=0.08) viscous fluid with Israel-Stewart's relaxation
equation Eq.\ref{eq4}, which neglect the term $R=[u^\mu\pi^{\nu\lambda}+u^\nu\pi^{\nu\lambda}]Du_\lambda$. The blue lines are energy density contours when the relaxation equation contain the term R. At early time, fluid evolution, with or without the term R remain essentially unchanged. Only at late time, evolution is marginally changed if the term is included in the relaxation equation.}
\end{minipage}
\end{figure}

 AZHYDRO-KOLKATA was tested extensively \cite{asis2nd} for numerical accuracy and consistency (see Fig.\ref{F1}).  
Many details of fluid evolution in minimally  viscous hydrodynamics can be found in \cite{asis2nd}. For completeness,  
in Fig.\ref{F2}, we have compared evolution of minimally viscous fluid and ideal fluid, in  a b=0 Au+Au collision, both initialised similarly. As expected, fluid cools slower with viscosity.  In Fig.\ref{F3}, evolution of shear stress tensor components, $\pi^{xx}$ and $\pi^{yy}$ in x-y plane is shown. Initially 
  $\pi^{xx}$ and $\pi^{yy}$ both have similar distribution. After a few fm of evolution, the distribution starts to differ. $\pi^{xx}$ and $\pi^{yy}$ are related by $x \rightarrow y$ and $y \rightarrow x$. The relation is maintained throughout the evolution. In Fig.\ref{F4}, black and red lines show the constant energy density contours in $r-\tau$ plane for ideal and viscous fluid respectively.  
In \cite{Romatschke:2007mq,Song:2007fn} it was observed that at late time, centre of the fluid cools faster in viscous evolution than in ideal evolution. However,
we do find a contrary result, even at late time, centre of the fluid cools slower than ideal fluid. 
However, in peripheral region and at late times, we do find that viscous fluid cools faster than ideal fluid. As mentioned earlier, relaxation equation Eq.\ref{eq3} neglect the term $R=[u^\mu\pi^{\nu\lambda}+u^\nu\pi^{\nu\lambda}]Du_\lambda$. To check that indeed
the term R contributes negligibly, in Fig.\ref{F4} we have shown results obtained with the term R included in the relaxation equation. 
The blue lines in Fig.\ref{F4} indicate that fluid evolution with or without the term R is nearly identical. For minimally viscous fluid neglect of the term    
  $R=[u^\mu\pi^{\nu\lambda}+u^\nu\pi^{\nu\lambda}]Du_\lambda$  in the relaxation equation is justified.

\section{Particle spectra and elliptic flow}

Hydrodynamic equations give the space-time evolution of thermodynamical quantities e.g. energy density (or temperature),   fluid velocity and the shear stress tensor. The information is converted into particle spectra using the standard Cooper-Frye prescription.  In Cooper-Frye prescription, invariant distribution of a particle, say $\pi^-$, is obtained as,

\begin{equation} \label{eq11}
\frac{dN}{dyd^2p_T} =\int_\Sigma d\Sigma_\mu p^\mu 
f^{(0)}(x,p)[1+\phi(x,p)] ,
\end{equation}

\noindent where $\Sigma_\mu$ is the freeze-out hyper surface and
$f^{(0)}(x,p)$ is the equilibrium one-body distribution function and
  $\phi(x,p)=\frac{\pi_{\mu\nu}p^\mu p^\nu}{2T^2(\varepsilon+p)}  << 1$, is the non-equilibrium correction to the equilibrium distribution function.   
Accordingly, in viscous dynamics,   invariant distribution has two parts,  

\begin{equation} \label{eq13}
\frac{dN}{dyd^2p_T}=
=\frac{dN^{eq}}{dyd^2p_T}+\frac{dN^{neq}}{dyd^2p_T},
\end{equation}

\noindent  where $\frac{dN^{neq}}{dyd^2p_T}$ is the non-equilibrium correction to the  
equilibrium distribution $\frac{dN^{eq}}{dyd^2p_T}$. Similarly, in viscous hydrodynamics,
elliptic flow of  can also be decomposed
in to two parts \cite{Teaney:2003kp},

\begin{equation} \label{eq14}
v_2(p_T)=\frac{\int d\phi \frac{d^2N}{dyd^2p_T} cos(2\phi)}
                {\int d\phi \frac{d^2N}{dyd^2p_T}  } \approx v_2^{eq}(p_T) + v_2^{neq}(p_T),
\end{equation}

\noindent  where $v_2^{eq}$ and $v_2^{neq}$ are the equilibrium flow (disregarding $\phi(x,p)$)   and its non-equilibrium correction respectively. 

\begin{figure}[t]
\begin{minipage}{12pc}
\includegraphics[width=12pc]{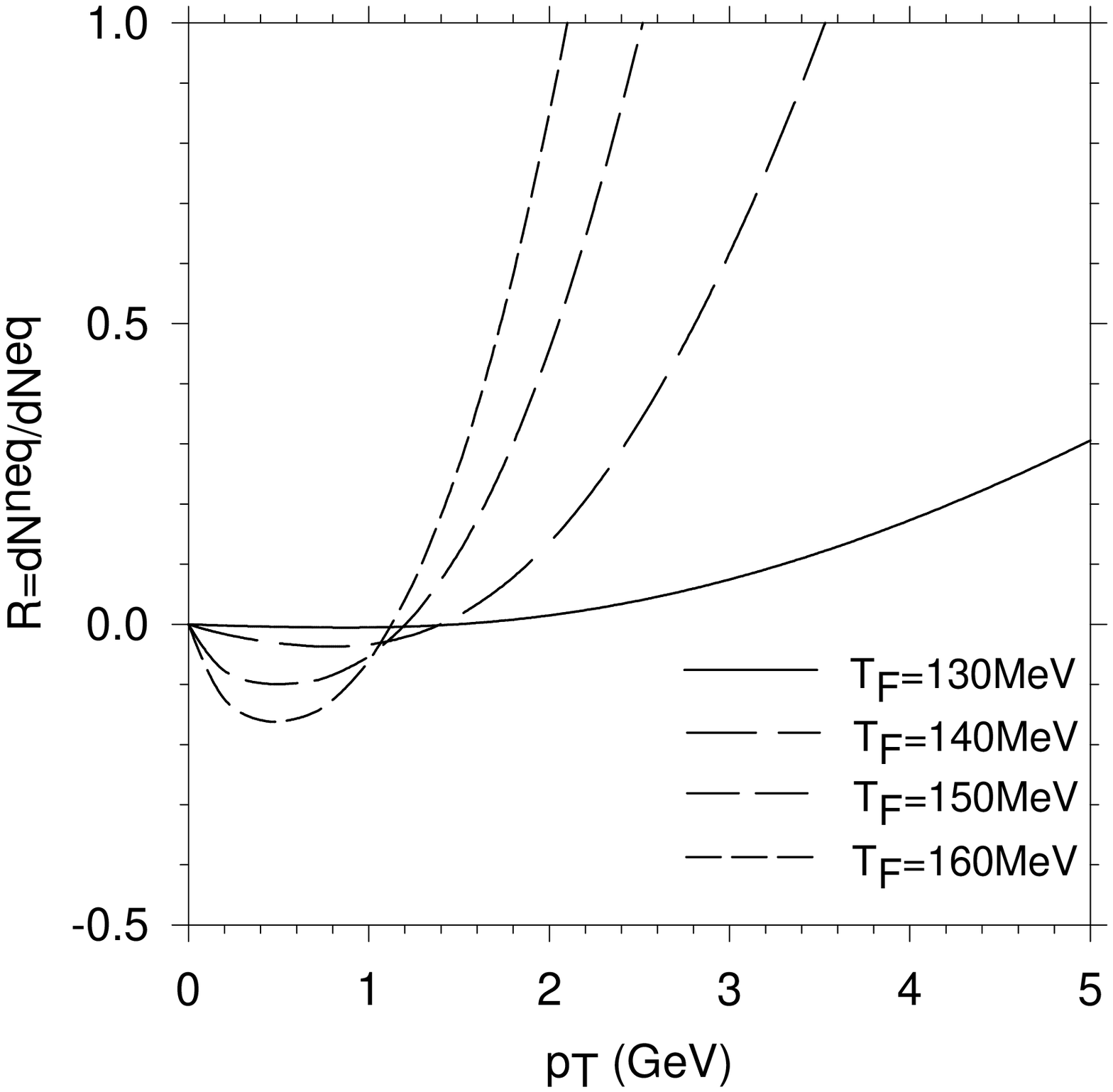}
\caption{\label{F6} Ratio of non-equilibrium to equilibrium contribution in b=6.5 fm Au+Au collisions.} 
\end{minipage}\hspace{1pc}%
\begin{minipage}{12pc}
\includegraphics[width=12pc]{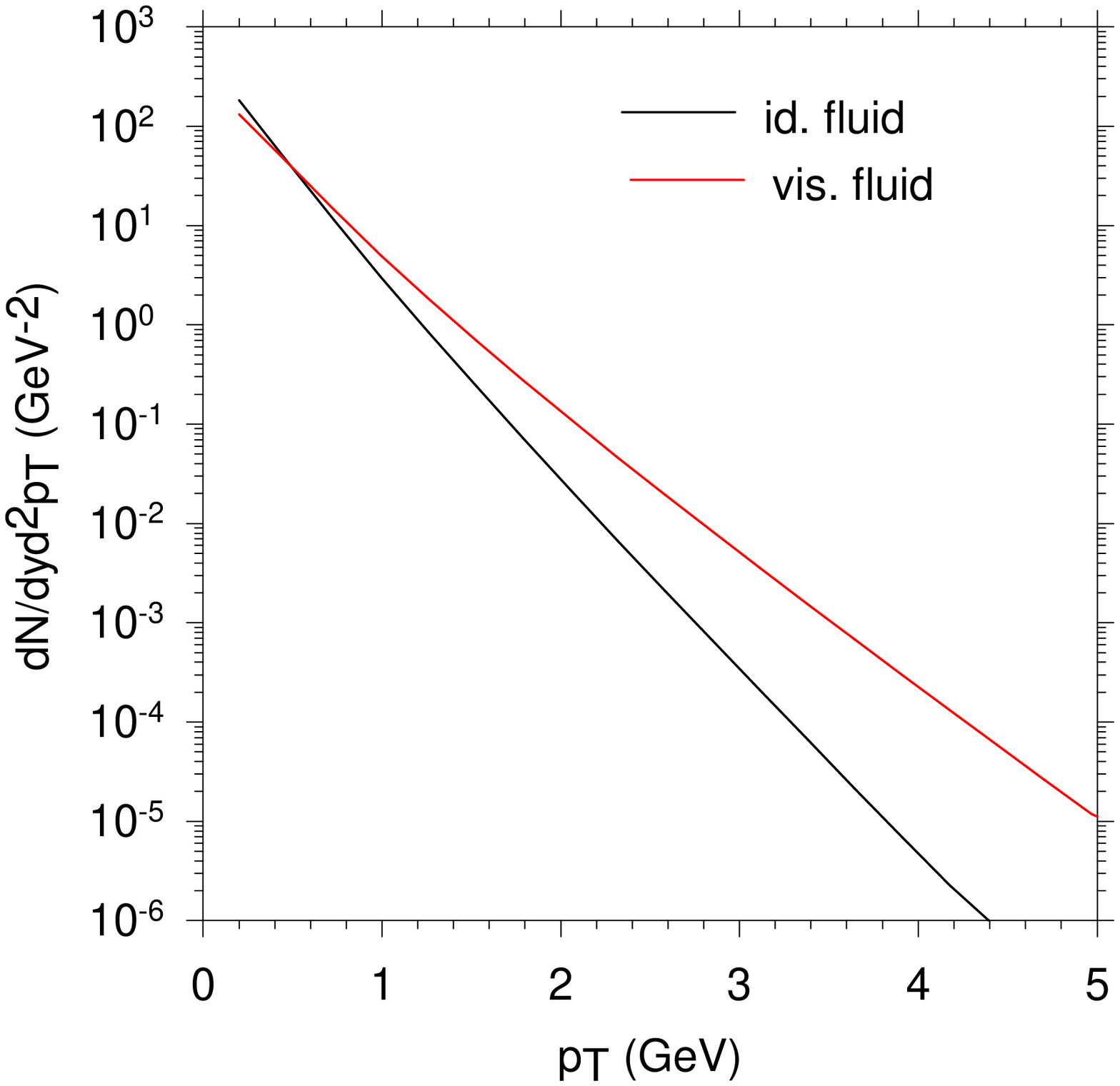}
\caption{\label{F7} $p_T$ spectra of $\pi^-$ in ideal and minimally viscous fluid, in b=7 fm Au+Au collisions.}  
\end{minipage}\hspace{1pc}%
\begin{minipage}{12pc}
\includegraphics[width=12pc]{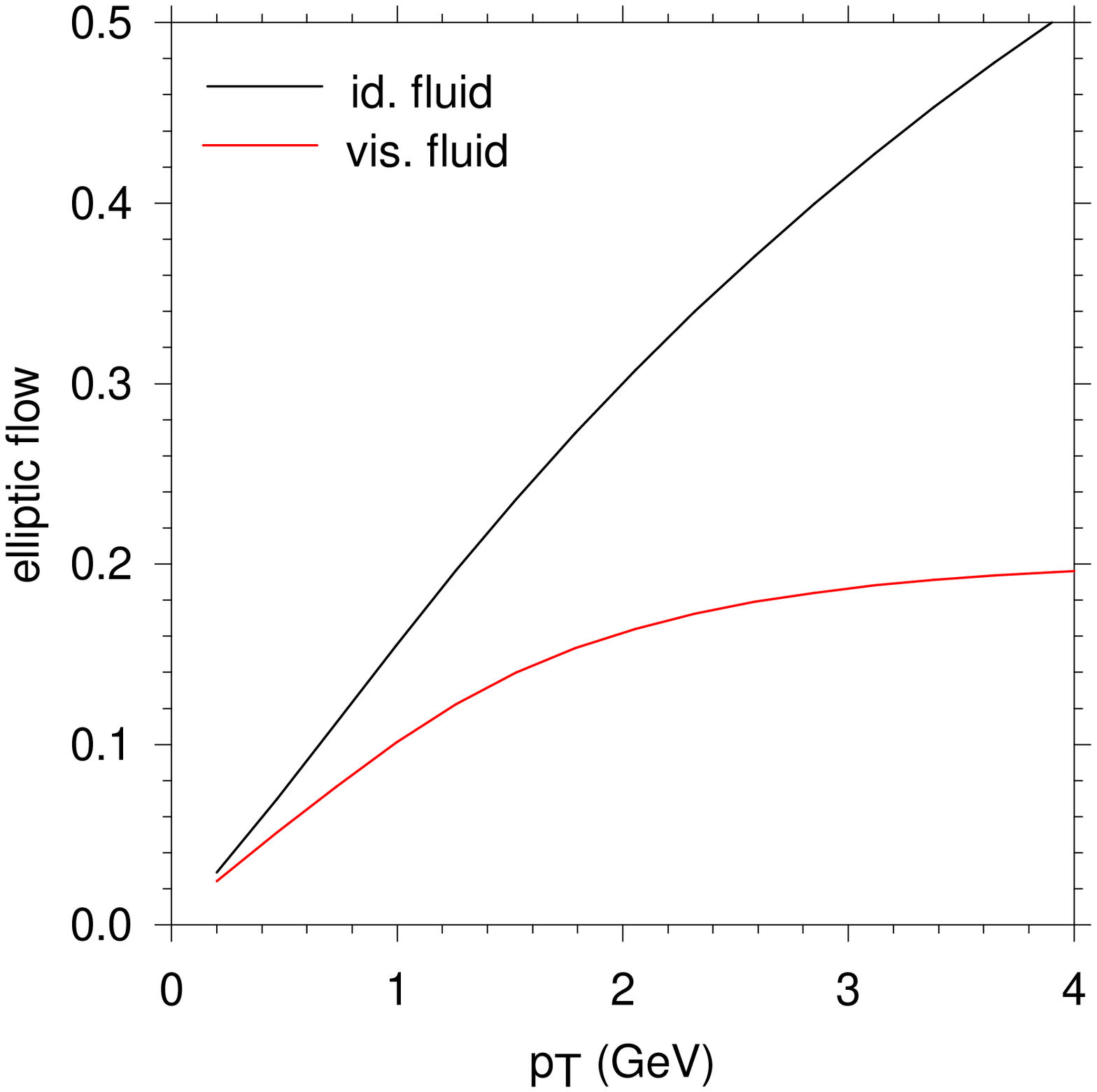}
\caption{\label{F8} Elliptic flow in ideal and minimally viscous fluid in b=7 fm Au+Au collisions.} 
\end{minipage}
\end{figure}

Since the non-equilibrium correction to equilibrium distribution $\phi(x,p) << 1$ it is then necessary that the ratio $R=\frac{dN^{neq}}{dyd^2p_T}/\frac{dN^{eq}}{dyd^2p_T} << 1$.
In Fig.\ref{F6}, $p_T$ dependence of the ratio $R$ in a b=6.5 fm Au+Au collision is shown. $R$ depend sensitively on the freeze-out temperature, decreasing with lowering $T_F$.   $p_T$ range over which hydrodynamics remain applicable also increases as $T_F$ is lowered. 
 
In Fig.\ref{F7} and \ref{F8}, we have demonstrated the effect of viscosity on $p_T$ spectra and elliptic flow. For both ideal and minimally viscous fluid, $\pi^-$ yield  in a b=7 fm Au+Au collision
is calculated from freeze-out surface at temperature $T_F$=130 MeV.  Compared to ideal dynamics particle yield is increased in viscous  dynamics, more at large $p_T$, e.g. at $p_T$=3 GeV, $\pi^-$ yield is increased by a factor of $\sim$10. Elliptic flow on the other hand decreases in viscous dynamics.  As shown in Fig.\ref{F8},   while  in ideal hydrodynamics elliptic flow continue to increase with $p_T$, it nearly saturate in viscous hydrodynamics.  
It is well known that ideal hydrodynamics under-predict the $p_T$ spectra and over-predict the elliptic flow. Viscous dynamics 
seems to remedy the drawbacks of ideal dynamics. 
 
\begin{figure}[ht]
\begin{minipage}{18pc}
\includegraphics[width=18pc]{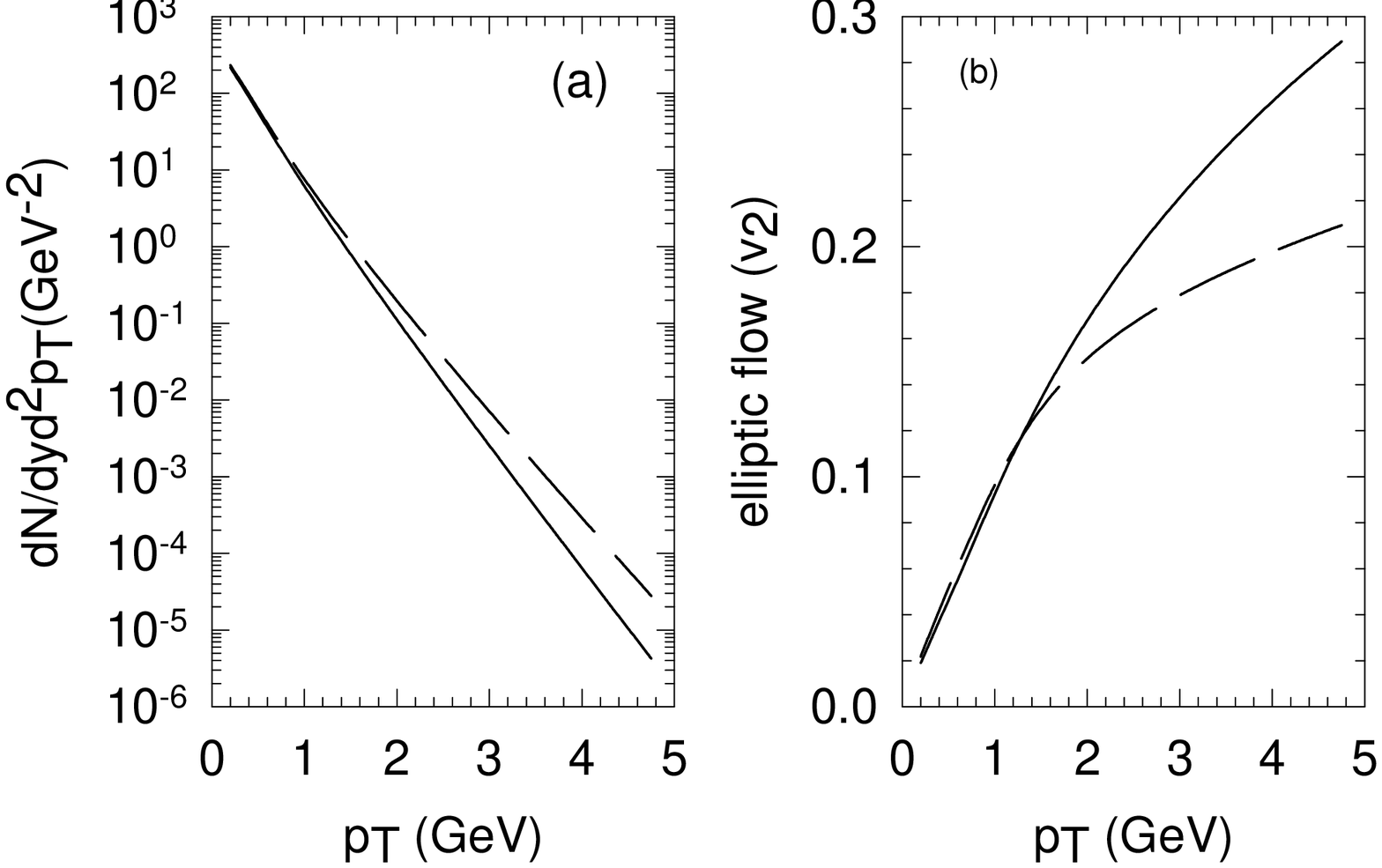}
\caption{\label{F15} (a) $p_T$ spectra and (b) elliptic flow in
b=6.5 fm Au+Au collisions.
Solid and dashed lines are for initial $\pi^{\mu\nu}=0$ and  $\pi^{\mu\nu}=2\eta/3\tau_i$ respectively.} 
\end{minipage}\hspace{2pc}%
\begin{minipage}{18pc}
\includegraphics[width=18pc]{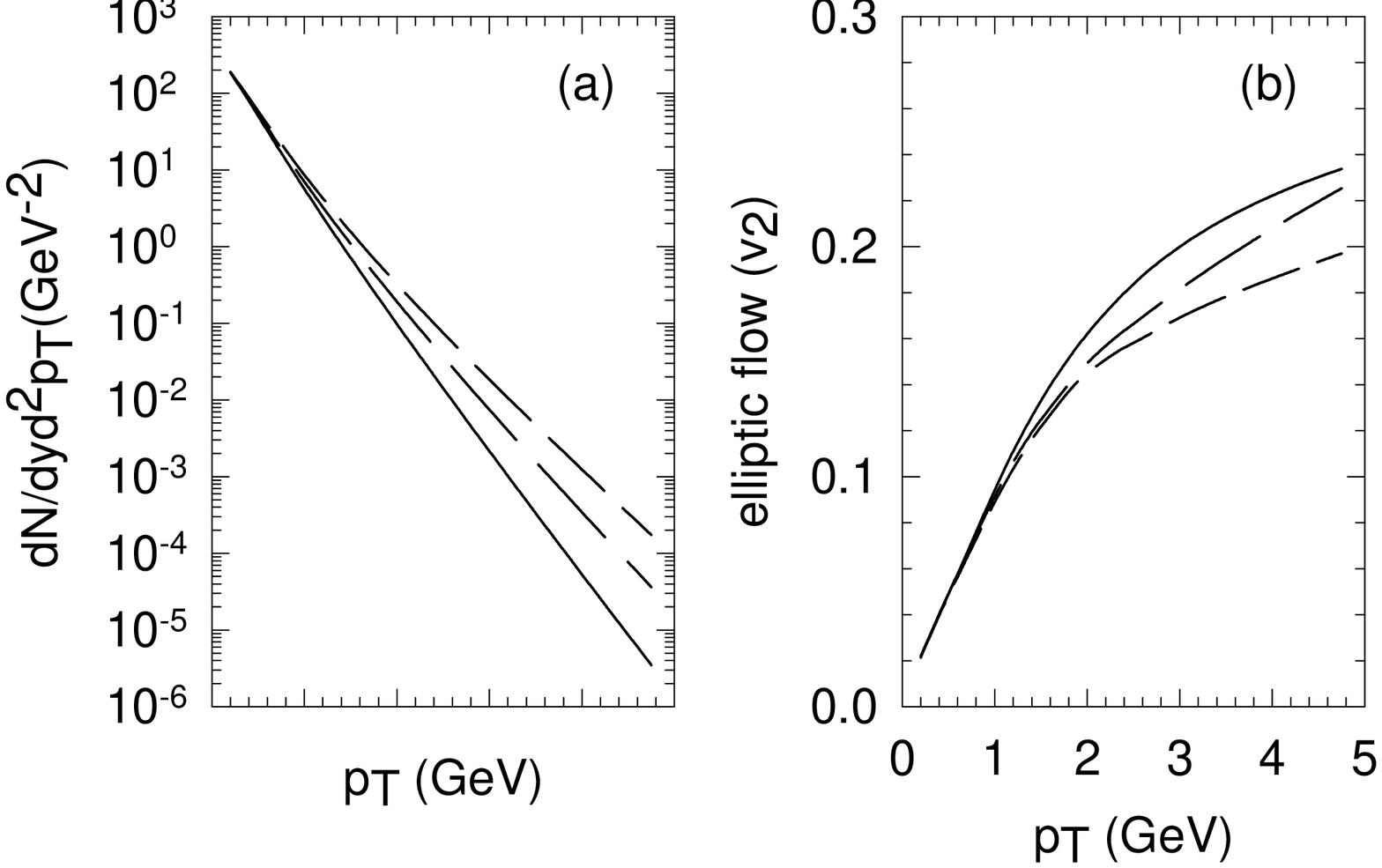}
\caption{\label{F16} (a) $p_T$ spectra and (b) elliptic flow in b=6.5 fm Au+Au collisions. The solid, dashed and short dashed liens are for $\tau_\pi$=$3\eta/sT$, $6\eta/sT$ and $9\eta/sT$ respectively. }  
\end{minipage} 
\end{figure}

\begin{figure}[t]
\begin{minipage}{16pc}
\includegraphics[width=16pc]{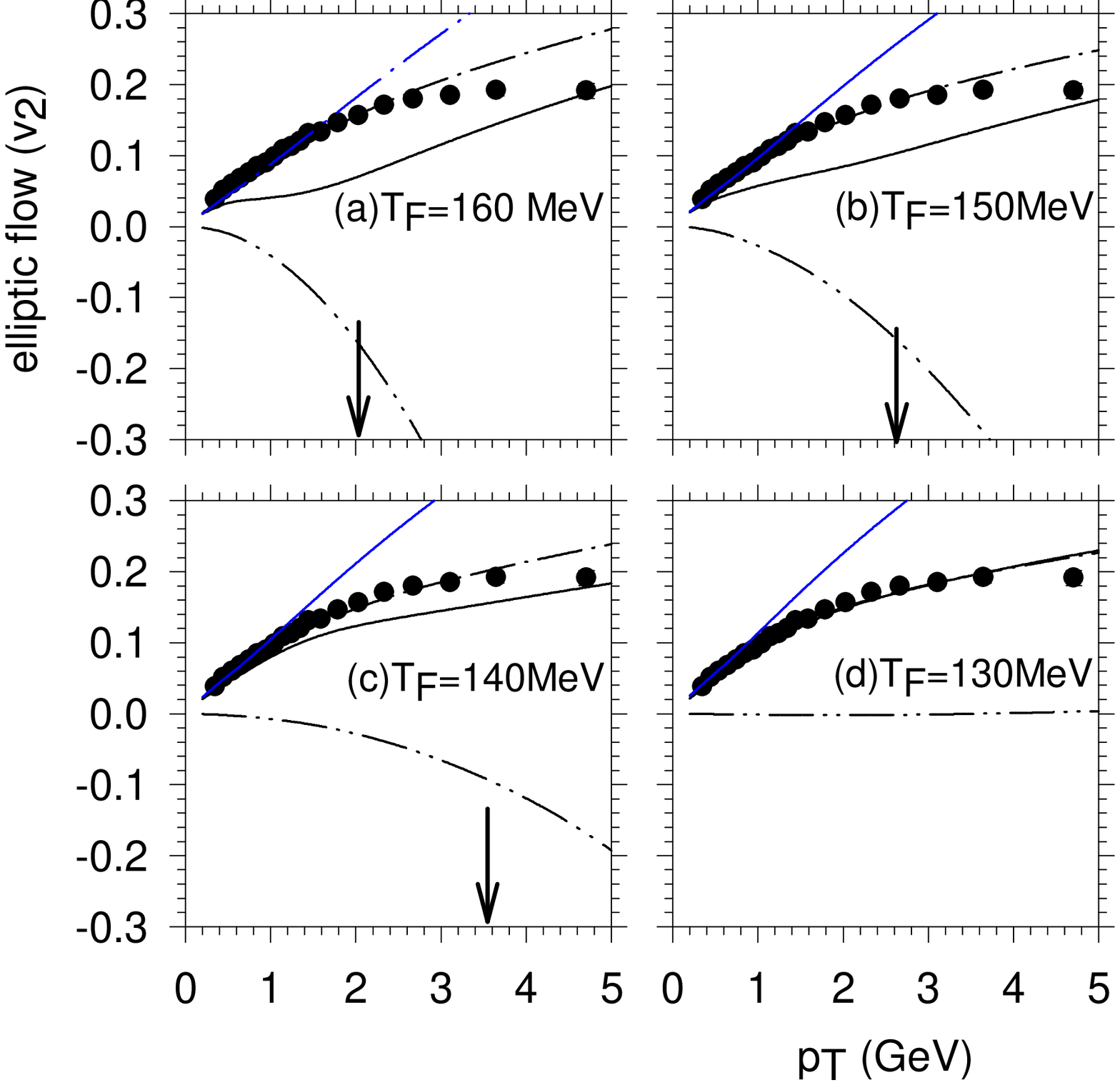}
\caption{\label{F9} PHENIX data on the $p_T$ dependence of elliptic flow in 16-23\% Au+Au centrality collisions are compared with minimally viscous fluid evolution (see text).}
\end{minipage}\hspace{4pc}%
\begin{minipage}{16pc}
\includegraphics[width=15.5pc]{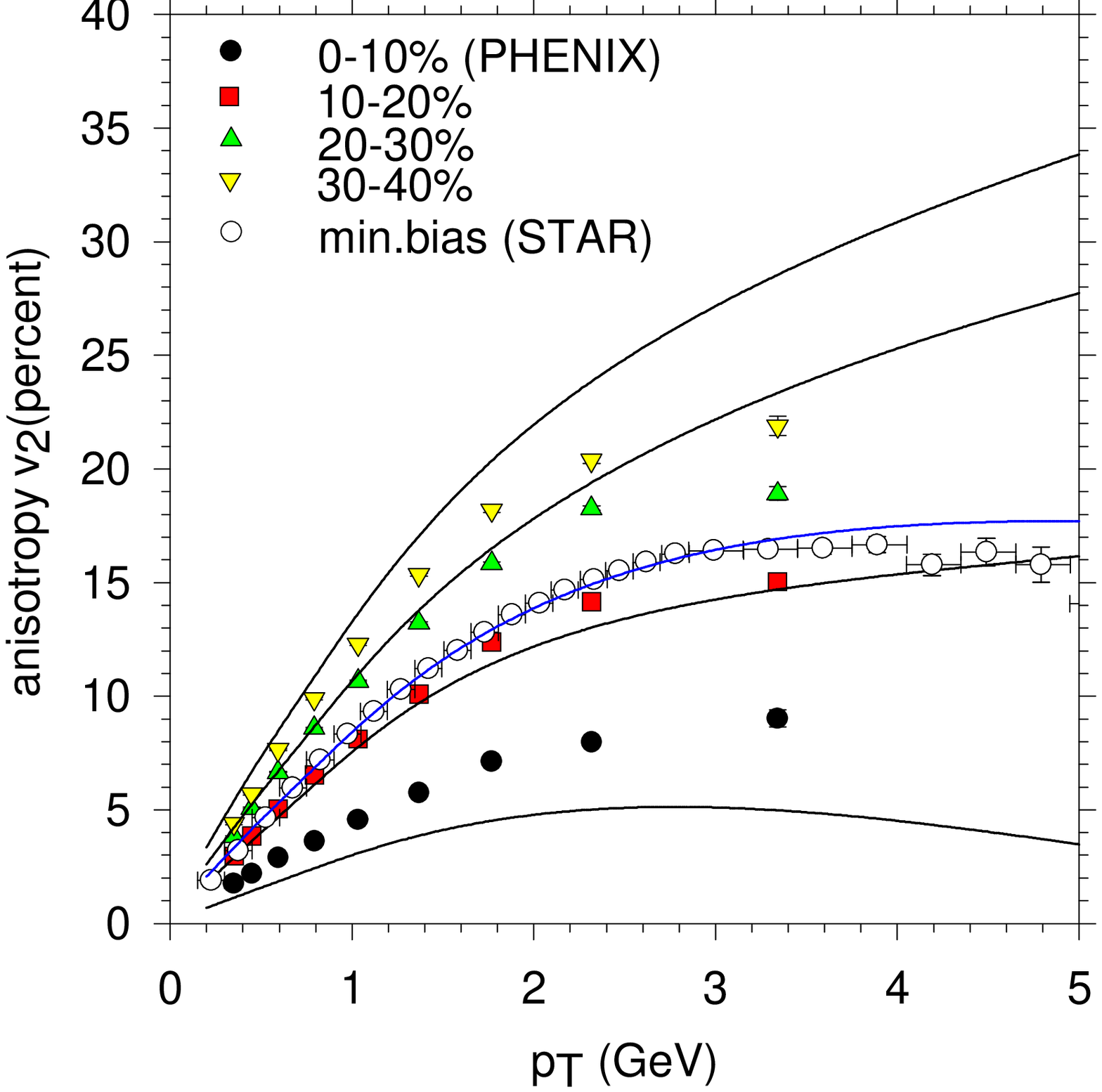}
\caption{\label{F10}Predictions from minimally viscous hydrodynamics for $v_2$ in different centrality ranges of collisions  are compared with the STAR and PHENIX data (see text). } 
\end{minipage}
\end{figure}

Data fitting is a complex process in viscous dynamics. Apart from initial time, initial energy density and fluid velocity (as in ideal hydrodynamics), 
in a minimally viscous fluid, particle production depends on the initial shear stress tensor, the relaxation time (see Fig.\ref{F15} and \ref{F16}). Viscous effects are enhanced  with increasing relaxation time.
Effects are also increased with non-zero initial $\pi^{\mu\nu}$, a result
contrary to \cite{Song:2007fn} where $p_T$ spectra or $v_2$ show little sensitivity to the initial shear stress tensor.  

Presently we fit the PHENIX data \cite{Adler:2004cj} on differential elliptic flow in 16-23\% centrality Au+Au collisions, by varying only $T_F$.  The other parameters are kept fixed, $\tau_i$=0.6 fm, $S_{ini}=110 fm^{-3}$, $v_x=v_y$=0, 
$\pi^{xx}=\pi^{yy}=2\eta/3\tau_i$,$\pi^{xy}=0$, $\tau_\pi=6\eta/sT$.
In Fig.\ref{F9}, elliptic flow in a b=6.5 fm Au+Au collision is compared with the PHENIX data for a range of freeze-out temperatures. We have shown the equilibrium flow $v_2^{eq}$ (the dash-dotted lines), the non-equilibrium correction $v_2^{neq}$ (the dash-dot-dotted lines) and total flow $v_2$  (the solid lines) separately.   It is interesting to note that $v^{eq}_2$ change  marginally from $T_F$=130-160 MeV .  It indicates that equilibrium flow is early time phenomena. Most of the flow is generated by the time fluid cools to T=160 MeV, and further evolution do not generate significant flow. Non-equilibrium correction is negative and as expected, reduces with lowering $T_F$. The black arrows in Fig.\ref{F9} indicate the $p_T$ above which viscous hydrodynamics break down. As seen in Fig.\ref{F9},
for $T_F$=130 MeV, the total elliptic flow agree well with the PHENIX experiment. For comparison, in Fig.\ref{F9} we have also shown the $v_2$ in ideal dynamics (the blue lines). In ideal dynamics, comparable fit is not   obtained.
     
As shown in Fig.\ref{F10} 
minimally viscous hydrodynamics, with $T_F$=130 MeV, also reasonably well explain the   PHENIX data \cite{Adare:2006ti} on differential $v_2$ in 0-10\%, 10-20\%, 20-30\% and 30-40\% centrality Au+Au collisions (the colored symbols) and the  
  STAR data \cite{Adams:2003zg}  (open symbols) for minimum bias $v_2$.
The black lines (from bottom to top) are $v_2$ from minimally viscous hydrodynamics in b=3.2, 5.7, 7.4 and 8.7 fm Au+Au collisions. They roughly corresponds to  0-10\%, 10-20\%, 20-30\% and 30-40\% centrality Au+Au collisions. $v_2$ in 10-20\% or in 20-30\% centrality collisions are reasonably well explained,   but $v_2$ is under predicted in 0-10\% centrality collision and over-predicted in 30-40\% centrality collisions.  
Interestingly,  STAR data on minimum bias $v_2$ is correctly explained  (the blue line in Fig.\ref{F10}). 
  In minimum bias, all the centrality ranges of collisions are included. 
Two opposing effects ($v_2$ under-predicted in central collisions and over predicted in peripheral collisions) are cancelled in minimum bias $v_2$. 
 
\begin{figure}[t]
\begin{minipage}{12pc}
\includegraphics[width=12pc]{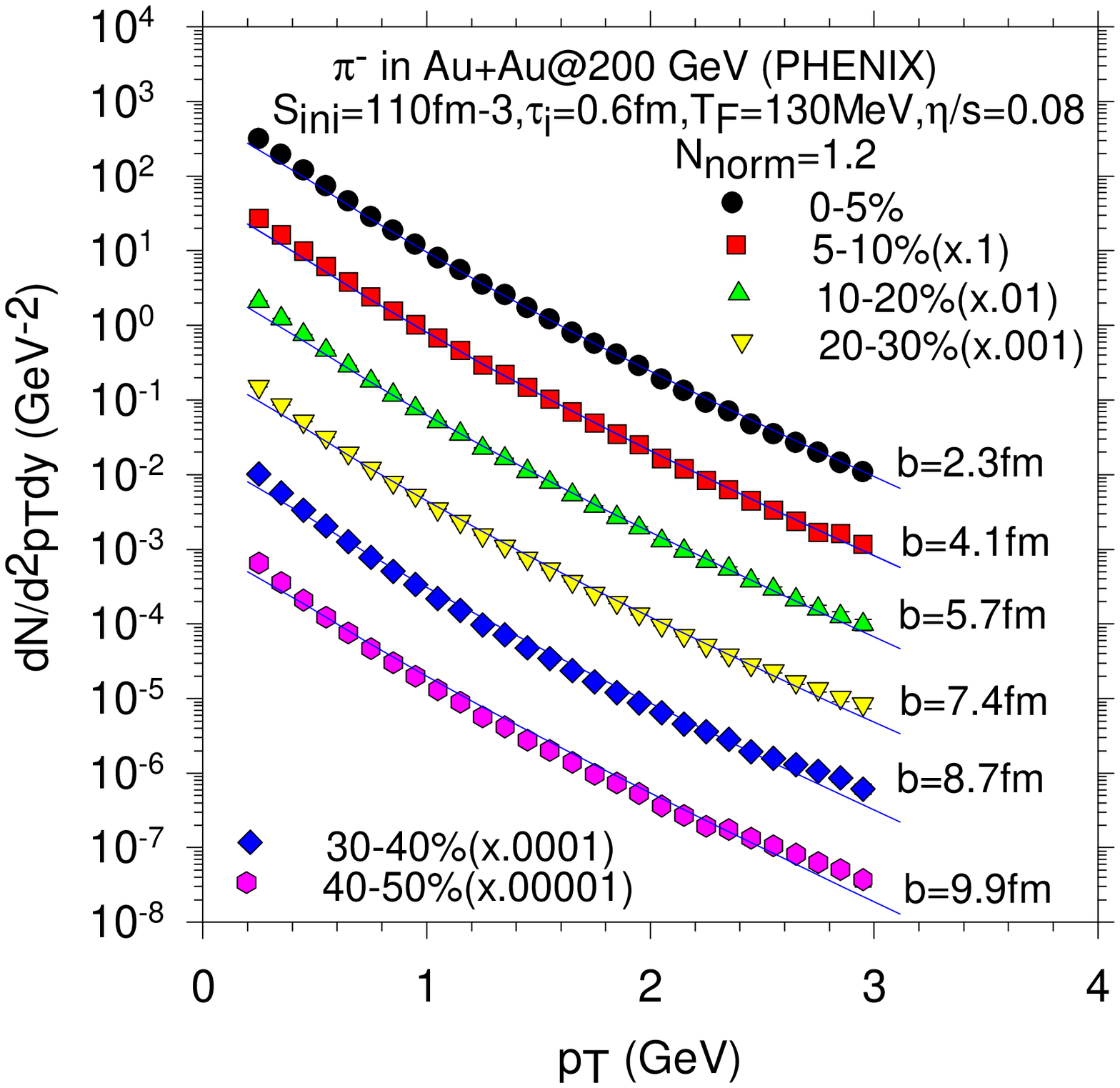}
\end{minipage}\hspace{1pc}%
\begin{minipage}{12pc}
\includegraphics[width=12pc]{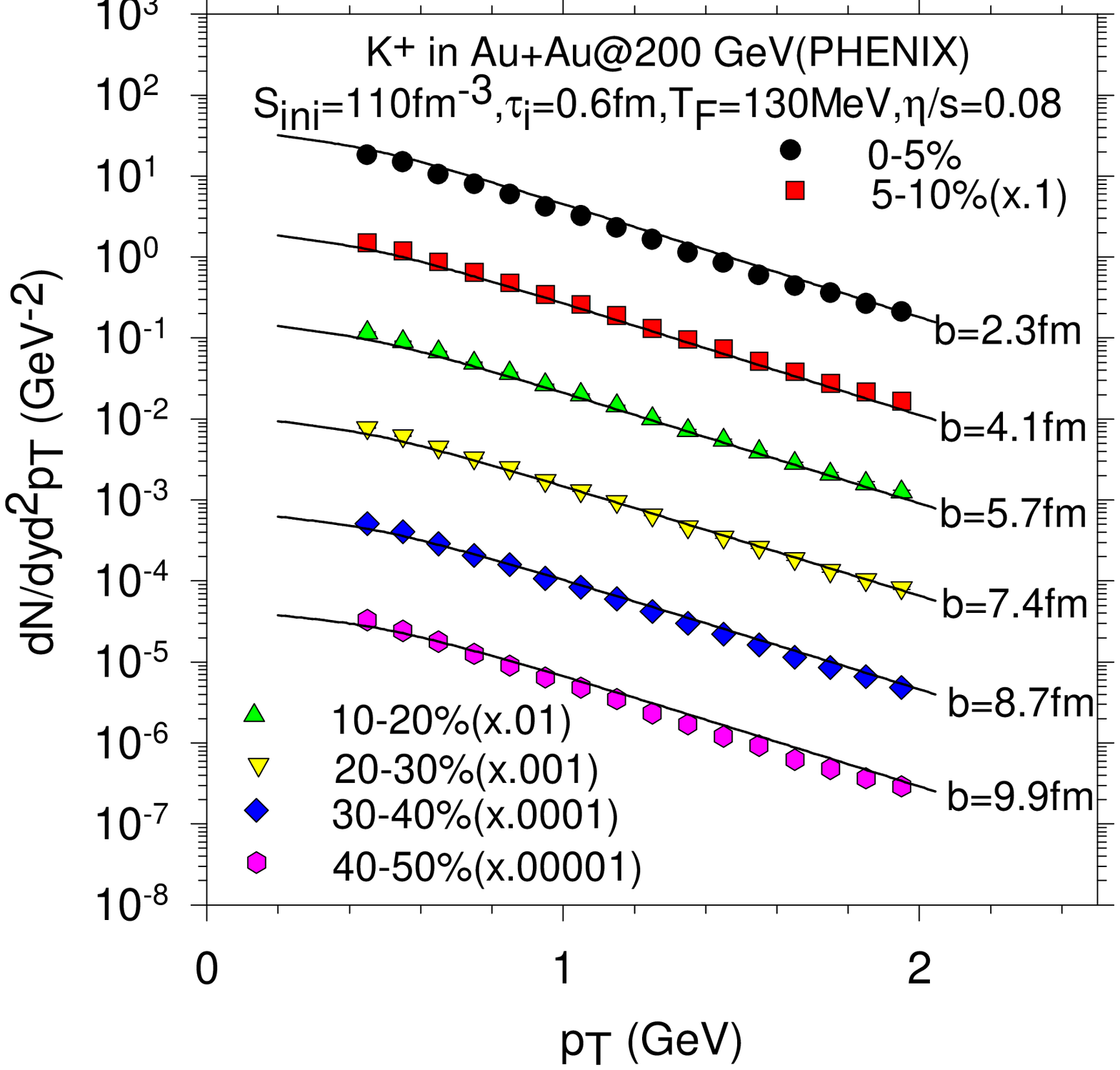}
\end{minipage}\hspace{1pc}%
\begin{minipage}{12pc}
\includegraphics[width=12pc]{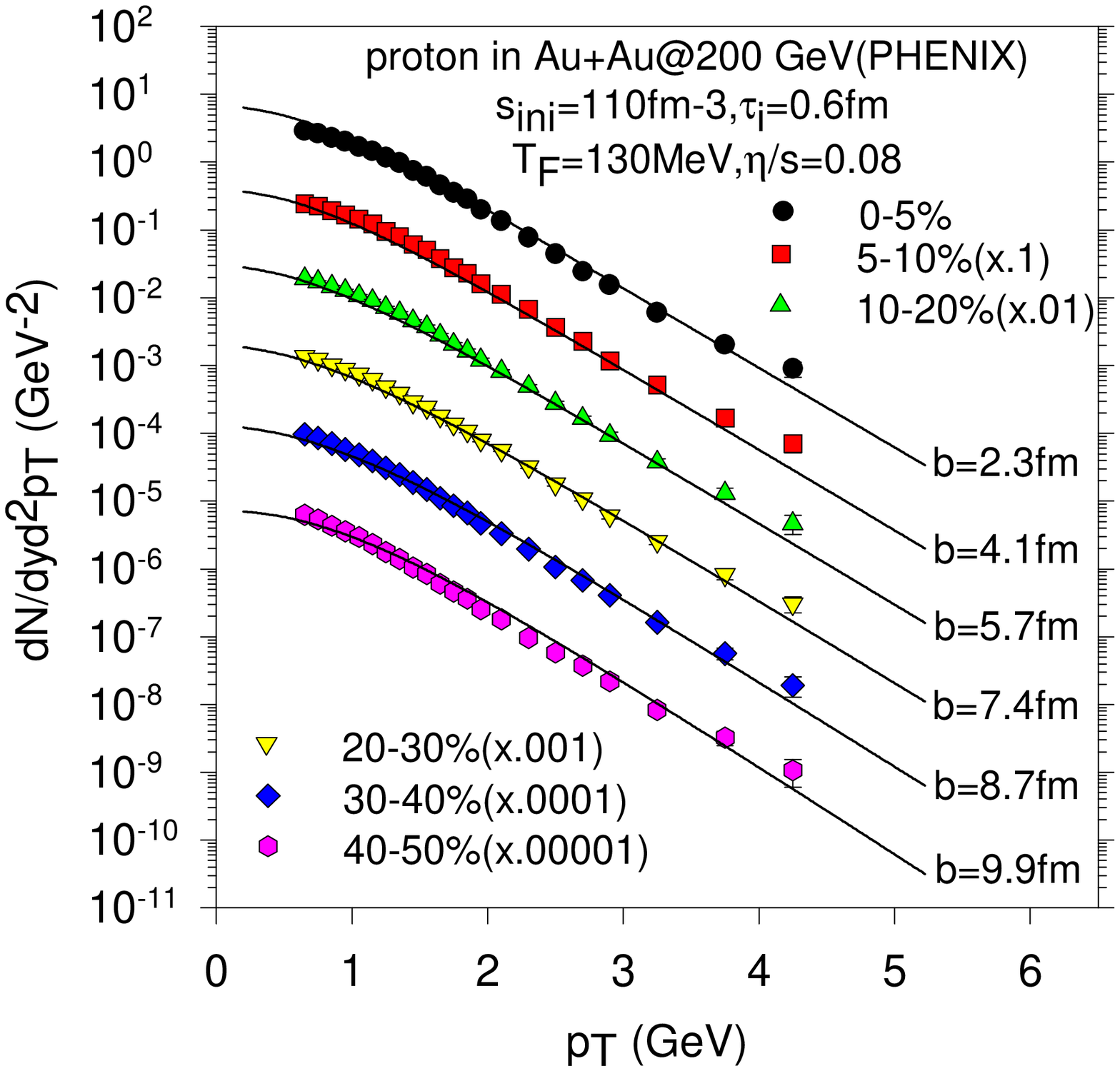}
\end{minipage}
\caption{\label{F11}(color online) PHENIX data \cite{Adler:2003cb} on $\pi^-$, $K^+$ and proton $p_T$ spectra,
in different centrality ranges of Au+Au collisions. Solid lines are 
predictions from minimally viscous hydrodynamics.} 
\end{figure}

\begin{figure}[t]
\includegraphics[width=16pc]{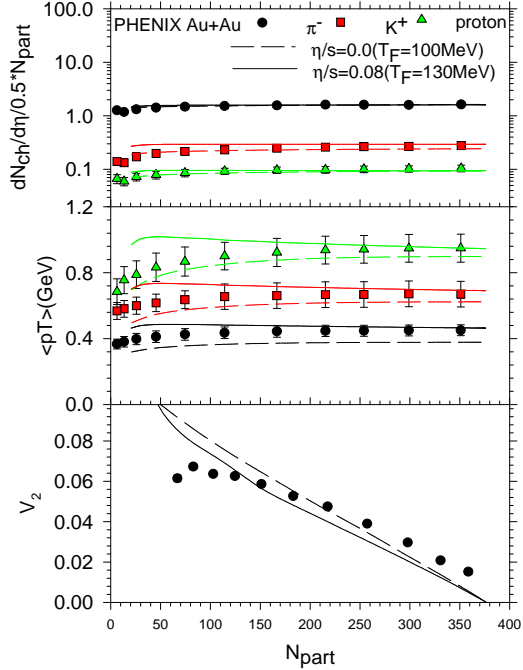}\hspace{2pc}%
\begin{minipage}[b]{16pc}\caption{\label{F14}
(color online) 
PHENIX data \cite{Adler:2003cb} on the centrality dependence of $dN_{ch}/dy$, $<p_T>$ and PHOBOS data \cite{Back:2004mh} on $p_T$-integrated $v_2$ are compared with predictions from minimally viscous dynamics (the solid lines) and ideal dynamics (the dashed lines). Freeze-out temperature is $T_F$=100 MeV for ideal and $T_F$=130 MeV for viscous fluid. Viscous hydrodynamics give nearly equivalent description of centrality dependence of $dN_{ch}/dy$ as in ideal dynamics. For $N_{part} \geq$100,  
centrality dependence of $<p_T>$ is better explained in viscous dynamics than in ideal hydrodynamics. PHOBOS data on integrated elliptic flow is also well explained.}
\end{minipage}
\end{figure}

Minimally viscous hydrodynamics with freeze-out temperature $T_F$=130 MeV also reproduces the centrality dependence of $p_T$ spectra  of  identified particles . In Fig.\ref{F11}  we have compared viscous hydrodynamics predictions with PHENIX data \cite{Adler:2003cb} on  $\pi^-$, $K^+$ and proton  $p_T$-spectra in 0-5\%, 5-10\%, 10-20\%, 20-30\%, 30-40\% and 40-50\% centrality Au+Au collisions. The spectra are 
are normalised by a factor $N=1.4$, which is reasonable considering that we have neglected resonance contribution. While ideal hydrodynamics can reproduce $p_T$ spectra only up to $p_T \leq 1.5$ GeV \cite{QGP3}, viscous hydrodynamics can reproduce the spectra throughout the $p_T$ range.  Viscous hydrodynamics 
also reproduces centrality dependence of charged particle multiplicity, mean  $p_T$ and $p_T$ integrated $v_2$ (see Fig.\ref{F14}).
 
The present study indicate that minimally viscous QGP fluid, with central energy density $\sim 35 GeV/fm^3$, thermalised at $\tau_i$=0.6 is consistent with experiments if the  hadronic fluid freeze-out at $T_F$=130 MeV. However, it must be mentioned that initial or final condition of the fluid may be changed if  a different $\tau_\pi$ or initial $\pi^{\mu\nu}$ is used. A comprehensive analysis,
exploring all the variables upon which the fluid evolution depend  is required to identify the initial condition of the fluid produced in Au+Au collisions at RHIC.
Apparently, present simulation  contradicts simulations in \cite{Romatschke:2007mq,Song:2007fn}. Different initial conditions, choice of system, evolution parameter etc. makes direct comparison between different simulations difficult. 
Song and Heinz \cite{Song:2007fn} studied viscous flow in Cu+Cu collisions.  They did not compare with data but $v_2$ in Cu+Cu indicate that possibly minimal viscosity will be inconsistent with the RHIC data. Romatschke et al   \cite{Romatschke:2007mq} studied Au+Au collisions. Minimum bias elliptic flow is under-predicted for minimal viscosity, data require $\eta/s \approx$0.03. However, they used  freeze-out temperature $T_F$=150 MeV and we expect they will find better agreement with data if the freeze-out temperature is lowered to $T_F$=130 MeV. 




\section{Summary}

To summarise, we have solved Israel-Stewart's 2nd order theory to study evolution of minimally viscous QGP fluid and subsequent particle production. In viscous dynamics, energy density or temperature of the fluid evolve slowly than in ideal fluid, for a fixed freeze-out temperature lifetime of the fluid is increased. Particle production is enhanced in viscous dynamics, more at large $p_T$. The elliptic flow on the other hand is reduced. Within certain approximation (non-zero initial shear stress tensor, kinetic theory approximation for relaxation time etc.),    ADS/CFT lower bound on viscosity appears to be consistent with a large number RHIC data in Au+Au collisions.    

\end{document}